\documentclass[aip,rsi,amsmath,amssymb,reprint,floatfix]{revtex4-1}
\usepackage{graphicx}
\usepackage{amsmath}
\usepackage{hyperref}
\usepackage{url}
\usepackage{dcolumn}
\usepackage{bm}
\usepackage[figure*,figure]{hypcap}

\begin{document}

\title{Scanning probe microscopy in an ultra-low vibration closed-cycle cryostat}%

\author{Francesca Paola Quacquarelli}
 \email[Francesca Paola Quacquarelli:]{francesca.quacquarelli@attocube.com}
\affiliation{attocube systems AG, K\"oniginstra\ss e 11a, 80539 M\"unchen, Germany.}
\author{Jorge Puebla}%
\affiliation{attocube systems AG, K\"oniginstra\ss e 11a, 80539 M\"unchen, Germany.}
\author{Thomas Scheler}%
\affiliation{attocube systems AG, K\"oniginstra\ss e 11a, 80539 M\"unchen, Germany.}
\author{Dieter Andres}%
\affiliation{attocube systems AG, K\"oniginstra\ss e 11a, 80539 M\"unchen, Germany.}
\author{Christoph B\"odefeld}%
\affiliation{attocube systems AG, K\"oniginstra\ss e 11a, 80539 M\"unchen, Germany.}
\author{Bal\'{a}zs Sipos}%
\affiliation{attocube systems AG, K\"oniginstra\ss e 11a, 80539 M\"unchen, Germany.}
\author{Claudio Dal Savio}%
\affiliation{attocube systems AG, K\"oniginstra\ss e 11a, 80539 M\"unchen, Germany.}
\author{Andreas Bauer}%
\affiliation{Physik-Department, Technische Universit$\ddot{a}$t M\"unchen, James-Franck-Str. 1, 85748 Garching, Germany.}
\author{Christian Pfleiderer}%
\affiliation{Physik-Department, Technische Universit$\ddot{a}$t M\"unchen, James-Franck-Str. 1, 85748 Garching, Germany.}
\author{Andreas Erb}%
\affiliation{Walther Meissner Institute, Bavarian Academy of Sciences and Humanities, 85748 Garching, Germany.}
\author{Khaled Karrai}%
\affiliation{attocube systems AG, K\"oniginstra\ss e 11a, 80539 M\"unchen, Germany.}

\date{\today}

\begin{abstract}
We report on state-of-the-art scanning probe microscopy measurements performed in a pulse tube based top-loading closed-cycle cryostat with a base temperature of 4$\,$K and a 9$\,$T magnet. We decoupled the sample space from the mechanical and acoustic noise from the cryocooling system to enable scanning probe experiments. The extremely low vibration amplitudes in our system enabled successful imaging of 0.39$\,$nm lattice steps on single crystalline SrTiO$_{3}$ as well as magnetic vortices in  Bi$_{2}$Sr$_{2}$CaCu$_{2}$O$_{8+x}$ superconductor. Fine control over sample temperature and applied magnetic field further enabled us to probe the helimagnetic and the skyrmion-lattice phases in Fe$_{0.5}$Co$_{0.5}$Si with unprecedented signal-to-noise ratio of 20:1. Finally, we demonstrate for the first time quartz-crystal tuning fork shear-force microscopy in a closed-cycle cryostat.\end{abstract}

\maketitle
\section{Introduction}
\label{sec:intro}
Magnetic force microscopy \cite{Hartmann1999} has recently allowed imaging of currents in topological insulators \cite{Nowack2013} and elusive nanoscale spin textures in chiral magnets, such as magnetic skyrmions\cite{Milde2013}. Similar fascinating many-body phenomena arise often at cryogenic temperature constraining scanning probe microscopy to be low-temperature compatible. State-of-the-art magnetic imaging, such as magnetic resonance force microscopy, with the detection of a single electron spin\cite{Rugar2004} and small nuclear spin ensembles\cite{Mamin2007}, are currently operated in liquid helium based cryostats. However, the high costs of helium and its scarcity have become a limiting factor. This situation has recently propelled the design and implementation of closed-cycle cryostats\cite{Radebaugh2009,Radebaugh1990,Chijioke2010,Uhlig2004} because they do not rely on the continuous supply of liquid helium. Still, the conjugation of scanning probe microscopy with the mechanically noisy environment of closed-cycle cryocoolers represents currently a challenge. A pioneering work on tuning fork-based scanning gate microscopy in a pulse tube cooled dilution refrigerator has been recently reported\cite{Pelliccione2013}. This achievement required a substantial redesigning of the cryocooling system and the introduction of damping measures on the microscope inside the cryostat, sacrificing thermal contact and requiring long pre-cooling times. In this work we report on the development of an ultra-low vibration closed-cycle cryostat enabling the top-loading of a variety of scanning probe microscopes for rapid experimental turnover. In particular, the microscopes do not require to be spring-mounted to further reduce vibrations, enabling the combination of high resolution confocal imaging and spectroscopy with scanning force microscopy. The ultra-low amplitude of the absolute vertical vibrations measured in our system enabled to clearly resolve, in AFM measurements, atomic steps of 0.39$\,$nm height in a SrTiO$_3$ sample. This level of vibrations has allowed the detection of the elusive skyrmion-lattice phase of a chiral magnet. It also opened the way to the implementation for the first time of tuning fork-based shear-force microscopy in a closed-cycle cryostat.
\section{Description of the cryocooling system}
\label{sec:system_description}
\begin{figure}[ht]
\centering
\includegraphics[width=\columnwidth]{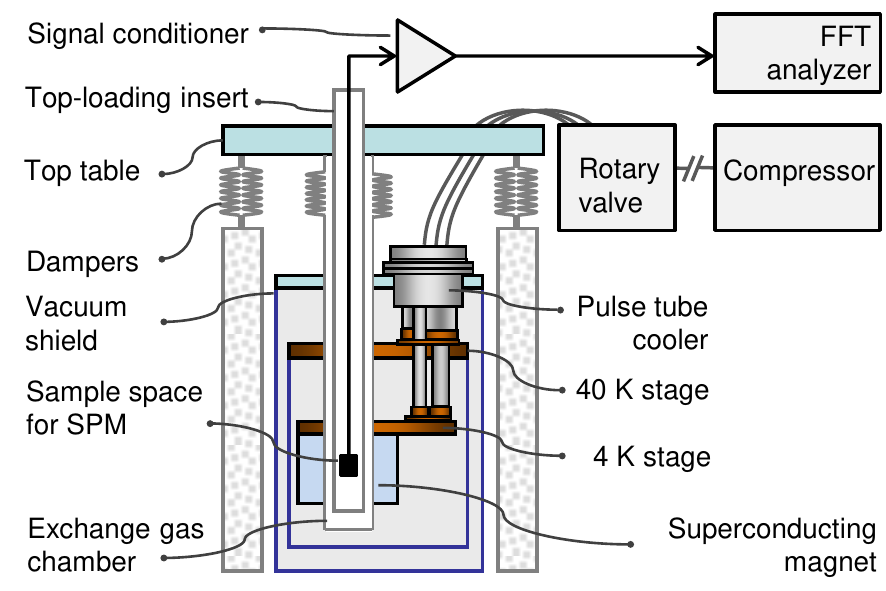}
\caption{Schematics of the closed-cycle cryocooling system based on pulse tube technology. The system consists of three main parts: a pulse tube, a rotary valve, and a compressor. A 9$\,$T superconducting magnet, fixed to the lower temperature stage of the pulse tube, surrounds coaxially the sample space and it is cooled by thermal contact. A vibration detector in its vacuum insert is top-loaded in the system at base temperature in order to characterize vibrations in the sample space.}
\label{fig:attodry_pic}
\end{figure}

The design of our  dry cryocooling system (see figure \ref{fig:attodry_pic}) is based on pulse tube technology\cite{Radebaugh2009,Wang1997,Radebaugh1990,Chijioke2010} and  relies on conductive cooling. In pulse tubes cryocoolers a cryogen fluid at high pressure is pulsed longitudinally  through the different stages of the tube at a frequency of 1.4$\,$Hz, reaching temperatures of approximately 3 K. Our system features two separate cooling stages at T = 40$\,$K (top stage) and T = 4$\,$K (lower stage). The cyclic expansion and compression of the gas in the adiabatic chamber where the fluid is stored determines a stratification of the temperature along the tube, so that one end is warmer than the other. The different pressure stages are separated by thermally conductive rare earths. The Cryomech PT-410 2-stage pulse tube cryocooler is driven by a compressor which provides 1$\,$W of cooling power at 4.2$\,$K. A rotary valve regulates the pressure of the gas through the pulse tube. A 9$\,$T superconducting magnet completes the system and is cooled by the pulse tube through thermal contact.
In our system's configuration, the cryostat and the rotary valve are located within the same room. The compressor is hosted in a room nearby: the 10-meter-long pipes through which the gas is pumped run along the floor and through an opening in the wall between the rooms. The ideal design of a closed-cycle cryostat (see figure \ref{fig:attodry_pic}) must provide mechanical decoupling between the pulse tube and the sample space in such a way that the vibrations from the pulse tube cryocooler do not influence vibration-sensitive experiments, while still ensuring a good thermal contact for sufficient cooling power. Initial cooling of the whole system requires 12 hours, including the magnet; cooling down of the microscope insert from T = 300 K to base temperature, namely 4$\,$K, is normally achieved within one hour, depending on the mass of the microscope. One hour is also required to warm the system up. Thanks to the top-loading architecture, the sample space is immediately accessible after warming up. The configuration of the system allows for measurements turnovers every three hours, including sample exchange times which typically require only a few minutes. The scanning probe microscopy experiments and the characterization of the vibrations in the system take place in the 50$\,$mm diameter sample space, which consists of an airtight removable thin wall stainless steel tubing, hosting the microscope insert. In figure \ref{fig:attodry_pic} the setup to measure the amplitude of the absolute vibrations is shown. It comprises an absolute vibration detector, a signal conditioner and a fast Fourier transform (FFT) analyzer. The vibration detector is hosted in the sample space while performing the tests. This detector is calibrated using interferometric methods. These results and the measurements of the absolute vibrations are discussed in the next section.

\section{Characterization of vibrations}
\label{sec:vibration detector}
\begin{figure}[ht]
\centering
\includegraphics[width=\columnwidth]{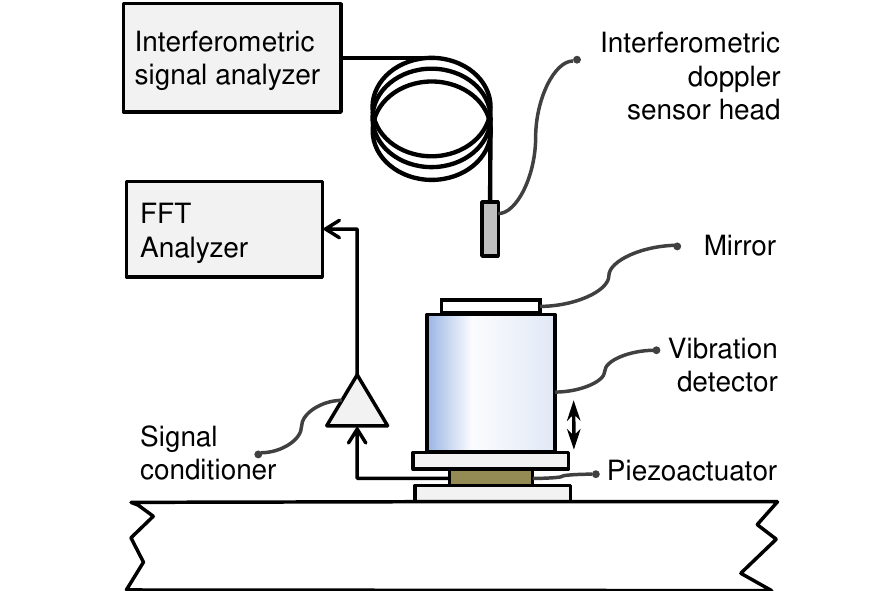}
\caption{Experimental setup for the characterization of the vibration detector in the vertical geometry by comparison with a doppler interferometer \cite{Karrai2007,Karrai2010,Thurner2013}. The vibration detector is mounted on the top of a plate driven by a piezoactuator. The arrow shows the direction of the piezo-induced vibration. A mirror on top of the vibration detector allows the detection of the displacement with the doppler interferometric sensor.}
\label{fig:vibration detector_FPS_setup}
\end{figure}

\begin{figure}[ht]
\centering
\includegraphics[width=\columnwidth]{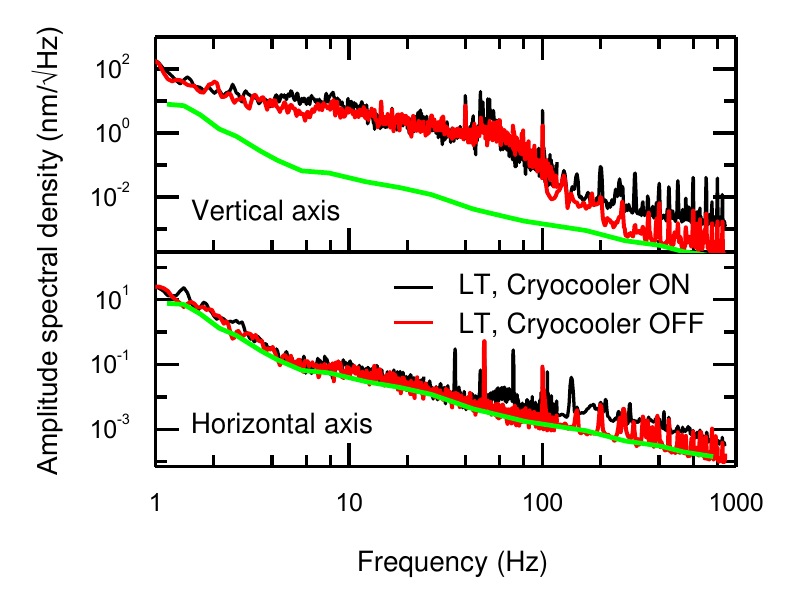}
\caption {Vibration amplitude measured in the cryostat at $T = 4\,K$ with the cryocooler ON (black line) and OFF (red line). The amplitude of the vibrations is reported in nanometers (nm) per square root Hertz. The top panel accounts for the vibration amplitude along the vertical axis; the bottom panel shows the amplitude of the vibrations along the horizontal axis. The green line represents the background level of vibrations measured independently in an isolated environment where we detect the lowest level of vibrations in our laboratory.}
\label{fig:vibration detector_vibrations}
\end{figure}

The amplitude of the vibrations in the system is measured using an absolute vibration detector, whose working principles are described in Ref.$\,$[\onlinecite{Stafford1953}]. We characterize the sensitivity of this vibration detector as a function of frequency using a doppler interferometer (attocube FPS3010 \cite{Karrai2007,Karrai2010,Thurner2013}) to measure the actual amplitudes. Figure \ref{fig:vibration detector_FPS_setup} shows the setup for the interferometric characterization of the vibration detector sensitivity, in the vertical direction.
\begin{figure}[ht]
\centering
\includegraphics[width=\columnwidth]{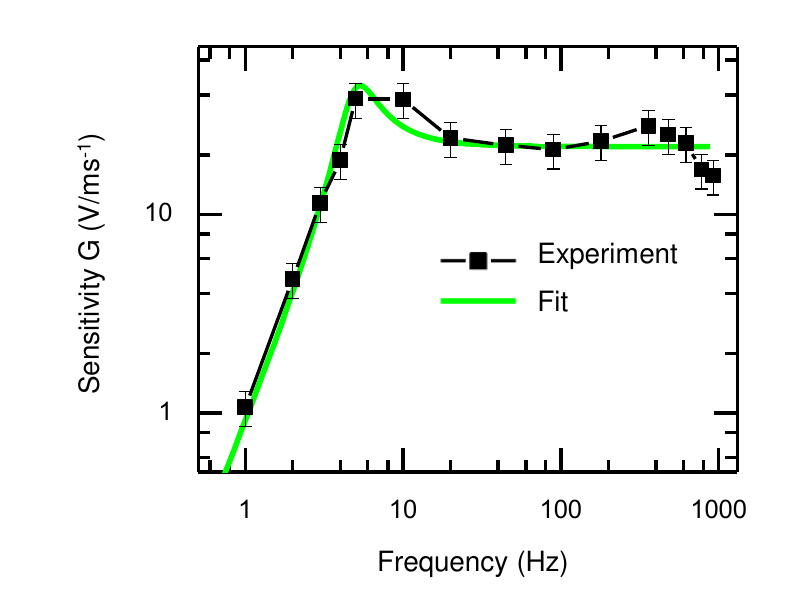}
\caption{The experimental curve of the sensitivity as a function of frequency of the vibration detector (solid squares) characterized using the interferometric sensor, reported in $V/ms^{-1}$. The green line shows a fitting curve of the data according to the dynamical equation of the velocity of a damped harmonic oscillator.}
\label{fig:vibration detector_FPS_graph}
\end{figure}
The vibration detector is mounted on a vibrating plate driven by a piezoactuator. In this experiment, a sinusoidal signal with a variable voltage is applied  to the piezoactuator to match at all times a constant amplitude of oscillation of 100$\,$nm detected with the interferometric sensor, while the frequency is varied within a range spanning from 1$\,$Hz to 1000$\,$Hz. We record the readings with the vibration detector and the results of the characterization of the sensitivity curve are shown in figure \ref{fig:vibration detector_FPS_graph}. The fitting curve is drawn according to the dynamical equation of the velocity of a damped harmonic oscillator. We experimentally determined the values for the calibration function G(f) according to the following expression:
\begin{equation}
    G(f) = \frac{V}{A\,2\pi\,f},
    \label{eq:sensitivity}
\end{equation}
where $A$ are the displacements readings given by the interferometric sensor, corresponding to the set peak-to-peak amplitude of 100 nm and $V$ represents the vibration detector read-out in peak-to-peak Volt. The measurements are reported in $V/ms^{-1}$ with error bars of  20$\%$. We used the experimentally determined calibration function G(f) to measure the vibrations in the sample space. Figure \ref{fig:vibration detector_vibrations} shows the spectral vibrational noise density along the vertical and the horizontal axis, measured in the sample space at T = 4$\,$K. The response of a vibration detector at low temperature was found consistent with its performance at room temperature. Figure \ref{fig:attodry_pic} shows the vibration detector setup for the characterization of vibrations in the sample space: the vibration detector is inserted in the vacuum tube at the location of the scanning probe microscope to simulate the experimental conditions. The two series of measurements for each direction of displacement correspond to acquisition with the cryocooling system on (black line) and off (red line). The green line represents the background level of vibrations measured independently in an isolated environment where we detect the lowest level of vibrations in our laboratory.
The data are shown in logarithmic scale to visualize the amplitude spectral density.
\begin{figure}[ht]
\centering
\includegraphics[width=\columnwidth]{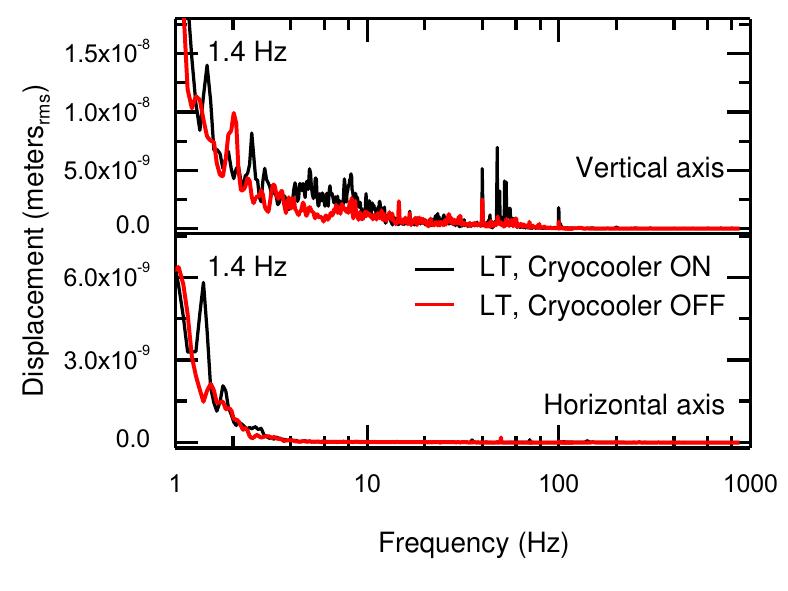}
\caption{Measurements of displacement determined by the vertical (top graph) and horizontal (bottom graph) residual vibrations in the sample space inside the closed-cycle cryostat. For each direction of displacement, the amplitude is measured at base temperature with the cryocooler ON (black line) and the cryocooler OFF (red line).}
\label{fig:vibrations_RMS}
\end{figure}
Figure \ref{fig:vibrations_RMS} shows the absolute vibration amplitude between 1$\,$Hz and 800$\,$Hz to reveal the technical spurious noise. The bandwidth used for the measurements was as follow for different intervals: 61$\,$mHz between 1$\,$Hz and 20$\,$Hz; 122$\,$mHz between 20$\,$Hz and 1000$\,$Hz; 1.9$\,$Hz between 100$\,$Hz and 800$\,$Hz. The rms vibration amplitude between 1$\,$Hz and 800$\,$Hz yields a peak value above background noise of 6 nm along the vertical axis and 3.5 nm (bandwidth 61$\,$mHz) along the horizontal axis at the pulse tube frequency f = 1.4$\,$Hz. The peak at the frequency of the gas pulse in the pulse tube is suppressed when the cryocooling system is switched off. The peaks at f = 50$\,$Hz and its harmonics are parasitic electromagnetic pick-up of the vibration detector and do not correspond to vibrations. This is verified by comparison between the measurements with the cryocooling system on and off and by performing the same measurements outside of the cryocooling system, in an isolated environment where we detect the lowest level of vibrations in our laboratory. We have demonstrated that the mechanical decoupling introduced here allowed to reduce the typical pulse tube generated vibrations by several orders of magnitude. The amplitude of vibrations is higher along the vertical axis since the pulses act along the vertical axis. The measured values of absolute vibration amplitude along the vertical axis in the nm range are low enough to enable scanning probe microscopy measurements as we characterize with contact mode AFM and prove in the following sections describing interesting magnetic force microscopy applications.
\section{Contact mode atomic force microscopy measurements}
\label{sec:AFM_noise}
\begin{figure}[ht]
\centering
\includegraphics[width=\columnwidth]{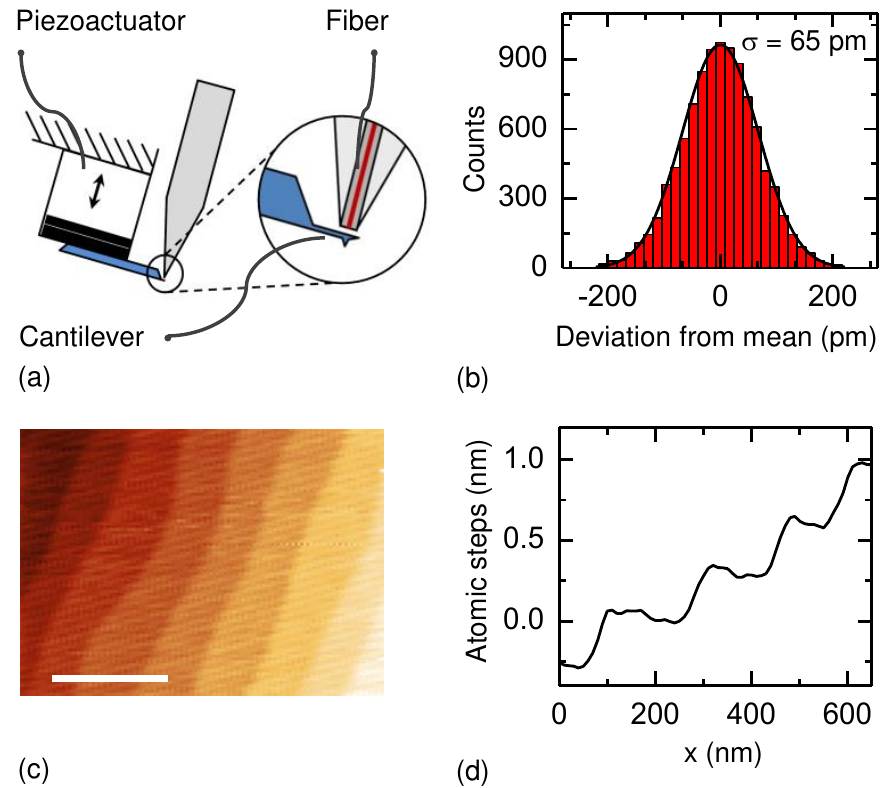}
\caption{(a) Interferometric head for AFM measurements. The deflection of the cantilever is detected with the built-in fiber-based interferometer. (b) Contact mode noise scan histogram of the z-height values measured with a bandwidth of 200 Hz at 3.2$\,$K. The measured amplitude of the relative displacement is 65 pm. (c) Contact-mode AFM image of atomically flat terraces on SrTiO$_{3}$ (200 scan lines) at 3.2$\,$K. The bar measures 400$\,$nm. The sample was prepared from a single crystal polished at 0.1$^{\circ}$ with respect to the crystal plane and subsequently annealed to ensure the formation of defined steps. The step height is 0.39$\,$nm, corresponding to the lattice parameter of the crystal. The frame time for the acquisition is 1680$\,$s at a scan rate of 500$\,$nm/s. (d) Single line profile showing the height of the atomically flat terraces on SrTiO$_{3}$.}
\label{fig:mfm_diagram}
\end{figure}

In contact mode atomic force microscopy (AFM) an etched silicon cantilever with a silicon tip combined with optical deflection detection is used to precisely measure local interactions such as van der Waals or Coulomb forces\cite{Giessibl2006}. The deflection detection in our system is performed by fiber-based interferometric sensing\cite{Rugar1989}: a fiber is mounted in the head of the microscope to detect the deflection of the cantilever induced by the tip-sample interaction. A measurement of the amplitude of the relative displacement between the AFM tip and the sample is performed to evaluate the impact of the absolute vibrations of the system on the performances of the scanning probe experiments. A schematic representation of our AFM head with interferometric read-out is shown in figure \ref{fig:mfm_diagram}a. The same configuration of the AFM head, with a magnetically coated cantilever and dither piezo excitation, is also used for the non-contact magnetic force microscopy\cite{Hartmann1999} experiments described in section \ref{sec:MFM}, where the cantilever is oscillated at a fixed distance from the sample.
We measure the z-noise while keeping the tip in contact with the sample surface. Vibration data describing the relative displacement between the tip and the sample for the fully enabled system are then acquired over time. The acquisition of noise statistics of the tip deflection is obtained over 10,000 points. The sampling time is set to 5$\,$ms, which corresponds to a measurement bandwidth of 200$\,$Hz. A noise histogram is shown in figure \ref{fig:mfm_diagram}(b). The result shows a normal gaussian distribution of the noise with a standard deviation $\sigma$ = 65$\,$pm rms; this value measured with the feedback loop enabled, compares to the z-noise amplitude obtained in liquid systems\cite{AG2013}. When the feedback is turned off, the noise amplitude increases approximately of a factor of 4. To demonstrate the low vibration noise we imaged atomically flat terraces of height corresponding to the lattice parameter a = 0.39$\,$nm in our dry cryostat at T = 3.2$\,$K on a strontium titanate (SrTiO$_{3}$) commercial wafer shallow polished at 0.1$^{\circ}$ then annealed to obtain terrace-and-step rearrangement of the outer layer of the crystalline structure. The image in figure \ref{fig:mfm_diagram}c shows a contact mode scan where the terrace-and-step morphology of the sample surface can be clearly observed, with steps measuring a single lattice constant. The image was acquired with 21$\,$ms sampling time, with a bandwidth of 47.6$\,$Hz. Our measurements are fully consistent with those reported of a similar commercial sample of SrTiO$_{3}$ measured at room temperature, showing 0.4$\,$nm lattice steps without the influence of a cryocooling system\cite{Rogalla2001,Huijben2009}. Such low level of vibrations opens the way to very sensitive scanning probe applications that we will now discuss.
\section{Magnetic force microscopy measurements}
\label{sec:MFM}
Physics at low temperature offers the challenging and rewarding opportunity of investigating fundamental properties of matter. In particular, the scanning probe family of magnetic imaging methods includes powerful techniques to probe fundamental interactions at the subatomic scale. Relevant research in solid-state physics is currently carried out by means of magnetic force and magnetic resonance force microscopy, imaging currents in topological insulators\cite{Nowack2013} and elusive nanoscale spin configurations (magnetic skyrmions) in chiral magnets\cite{Milde2013}, scanning diamond magnetometry, using nitrogen-vacancy (NV) centers in diamonds as solid-state sensors of magnetic field\cite{Wrachtrup2008,Hong2013,Mamin2013}, and scanning gate microscopy\cite{Pelliccione2013,Garcia2013}. These techniques enable sensitivity to extremely small magnetic interactions with nanometer resolution and have been recently successfully implemented in liquid bath cryostats. Given the scarcity of liquid helium, combining scanning magnetic force microscopy measurements with dry cryostats represents the way forward. Thanks to the low level of vibrations achieved in our dry cryostat we were able to carry out a variety of magnetic force microscopy measurements, whose description we report in this section. Magnetic force microscopy\cite{Hartmann1999,Giessibl2009} (MFM) is a technique derived from atomic force microscopy\cite{Binnig1986,Giessibl2003,Giessibl2006}; it takes advantage of tips with magnetic coatings, typically NiCr or cobalt (Co), making them sensitive to the variations of magnetic field. The strength of the magnetic interaction between the tip and the magnetic fields near the surface determines the vertical displacement of the tip while it is scanned across the sample. Figure \ref{fig:mfm_diagram}a in the previous section shows a schematic of the cantilever-based force microscope used in this work, designed particularly for applications at low temperature and high magnetic field. The tip deflection is detected using a single-mode, fiber-based interferometer\cite{Rugar1989}. As with most microscopes of the same kind, our magnetic force microscope applies an AC modulation technique to achieve highest detection sensitivity\cite{Zech2011}. Probes and the dynamic modes of operation are described in Ref.$\,$[\onlinecite{Hartmann1999}]. Magnetic force microscopy is especially sensitive to the sample-tip separation and vertical vibrations can affect the resolution of the image by impairing the force sensitivity of the probe and drastically reducing the resolution of the image. We measured magnetic vortices and elusive skyrmion-lattice phase in our dry cryostat with unprecedented signal-to-noise ratio.
\subsection{Magnetic vortices}
\label{subsec:magnetic_vortices}
\begin{figure}[ht]
\centering
\includegraphics[width=\columnwidth]{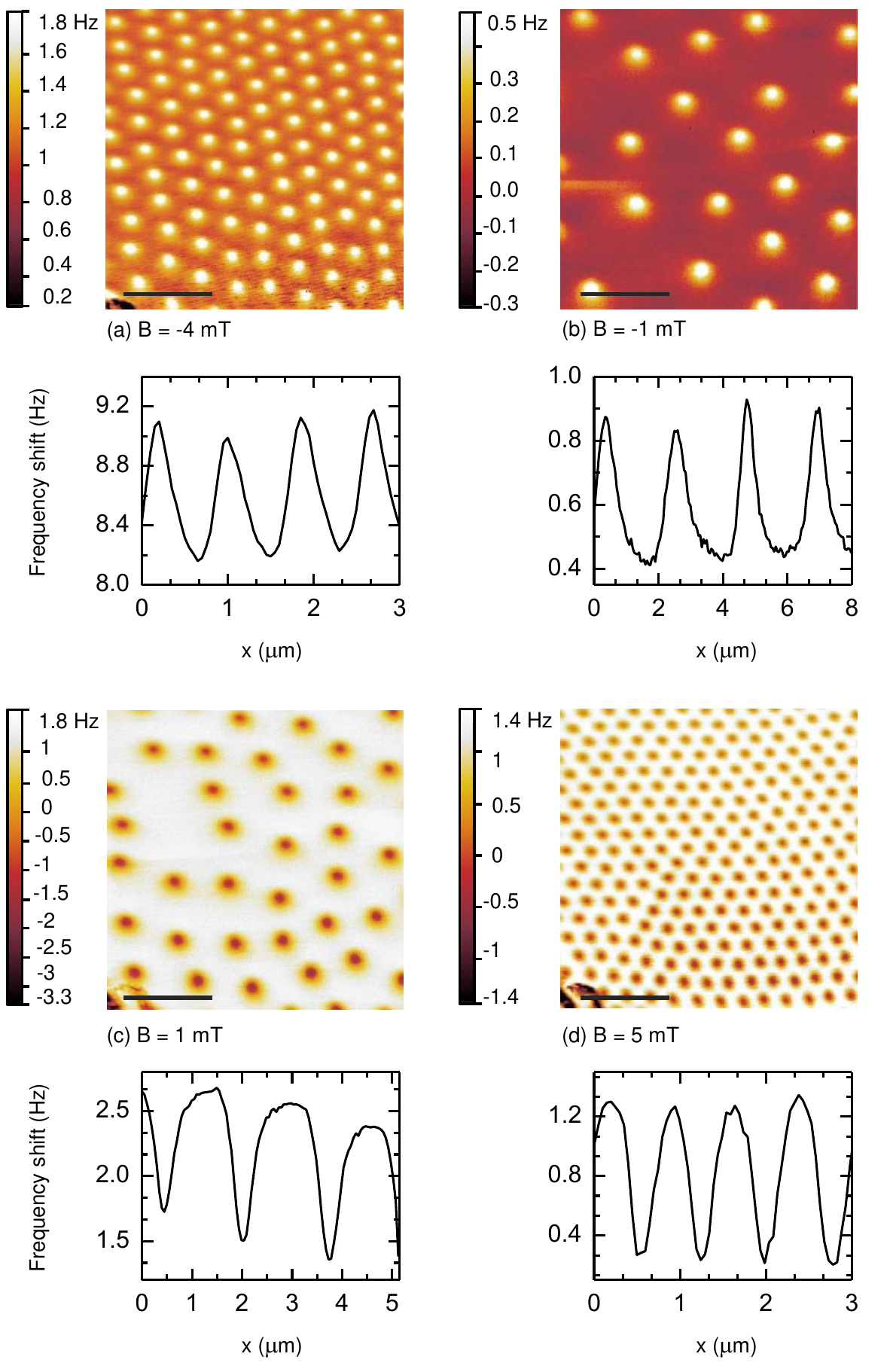}
\caption{MFM measurements of magnetic vortices lattices at different magnetic fields, from (a) to (d): B=-4$\,$mT, -1$\,$mT, 1$\,$mT, 5$\,$mT. The image shows 10 $\times$ 10$\,\mu m^{2}$ scans (200 by 200 lines) performed in constant height mode, around 30 nm above the sample, with an active phase-locked loop (PLL) at T = 4$\,$K. The bar in the images measures 2$\,\mu m$. Line cuts of the sample highlight the frequency shift.}
\label{fig:mfm_images}
\end{figure}

Vortices in superconductors of type II are interesting for both the theoretical challenge in unraveling the nature of high-T$_{c}$ superconductivity \cite{Bedorz1986} and their practical applications in material optimization for high current transport. The magnetic properties of a superconducting vortex originate in a circular supercurrent, which allows one magnetic flux quantum $\Phi_{0}=2.07\times10^{-15}\,$Tm$^{2}$ to penetrate the superconductor. The density of vortices can be tuned and it is directly proportional to the external applied magnetic field in the intermediate region between the lower critical magnetic field $H_{c1}$ and the critical field $H_{c2}$ above which superconductivity vanishes. Vortex lattices form as the result of minimization of the free energy density, assuming complex crystal structures, see e. g. Ref.$\,$[\onlinecite{Cubitt1993,Riseman1998,Bianchi2008,Mddotuhlbauer2009}], in the absence of local pinning forces. Pinning can be either artificial or intrinsic to the bulk of the superconductor. In the presence of such forces, originating from defects in the superconducting material, vortices are immobilized at pinning sites. While vortices can move upon application of an external current and induce electrical resistance, controlled tailored pinning is desirable to reduce this adverse effect and minimize electrical losses.  We investigated single magnetic vortices in freshly cleaved Bi$_{2}$Sr$_{2}$CaCu$_{2}$O$_{8+x}$ (BSCCO-2212), a cuprate, high-temperature superconductor\cite{Yuan1996}. The measurements in figure \ref{fig:mfm_images} show how the density of vortices varies with applied magnetic field when varying it from -4$\,$mT to 5$\,$mT.
The orientation of the vortices with respect to the moment of the tip is indicated by the colour of the vortices: bright (dark) colours indicate repulsive (attractive) forces. Here, the tip was scanned at the constant height of about 30$\,$nm above the surface of a freshly cleaved piece of BSCCO-2212 (sample courtesy of A. Erb, TU Munich). Note that the applied field was always much lower than the coercitivity of the hard-magnetic tip ($\sim$ 40$\,$mT), hence the orientation of the tip moment was kept unchanged. While at low vortex densities (figures \ref{fig:mfm_images}(b) and (c)) pinning effects are comparable and the lattice is manifestly disordered, one can see at higher densities the ordered hexagonal Abrikosov pattern\cite{Abrikosov1957}. Different compounds either iron or copper based exhibit superficial or bulk pinning  upon the effect of different critical currents\cite{Yin2009}. Accurate tailoring of pinning is most desirable for wiring for superconducting ultra-high-field magnets or power grids. As an indicator for the performance of the system, we measured the signal-to-noise ratio (SNR) from the peak heights of the isolated vortices for example at B = 1$\,$mT to be higher than 20:1, with a 10$\,$ms bandwidth, matching the SNR measured in low noise liquid helium cryostats \cite{Volodin1998}.
\subsection{Skyrmion-lattice and helimagnetic phase in single-crystal Fe$_{0.5}$Co$_{0.5}$Si}
\label{subsec:skyrmions}
\begin{figure}[ht]
\centering
\includegraphics[width=\columnwidth]{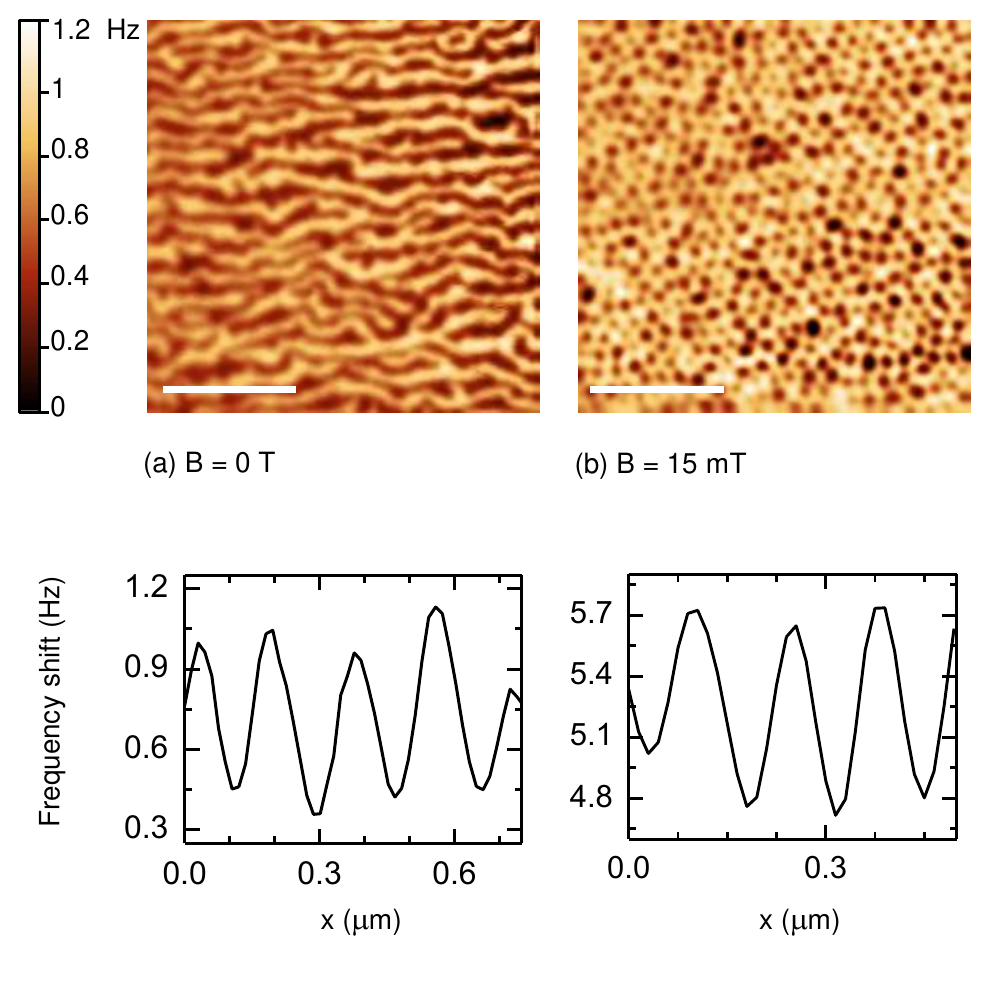}
\caption{MFM images of a polished surface of bulk samples of Fe$_{0.5}$Co$_{0.5}$Si. The images consist of 200 scan lines and were acquired using a sharp commercial cantilever (tip apex radius $\sim$ 10$\,$nm, SSS-MFMR from Nanosensors) with magnetic coating. (a) Helimagnetic phase of the sample at T = 3.2$\,$K after zero-field cooling (B = 0$\,$T). The difference in contrast from left to right is produced by the slope correction settings. (b) Metastable skyrmion-lattice phase measured at T = 3.4$\,$K in an external magnetic field B = 15$\,$mT after field-cooling. Bars in the images correspond to 1$\,\mu m$. Line cuts highlight the frequency shift.}
\label{fig:chiral_mag}
\end{figure}

Magnetic skyrmions are nanoscale spin textures in chiral magnets. The name skyrmions allude to a non-linear field theory proposed in the context of nuclear physics by T. H. R. Skyrme\cite{Skyrme1961}. In magnetic materials corresponding topologically nontrivial configurations were recently discovered within a narrow region of temperature and magnetic field\cite{Yu2010}. The skyrmion-lattice phase can be achieved in materials without inversion symmetry\cite{Bak1980}. The topological stability of magnetic skyrmions makes them excellent candidates for magnetic storage\cite{Plfleiderer2010,Nagaosa2013}. Skyrmions as little as 1$\,$nm were recently reported \cite{Heinze2011}. Particularly appealing is the observation of extremely efficient spin transfer torques at ultra-low current densities, roughly six orders of magnitude smaller than anything observed in conventional ferromagnetic materials\cite{Jonietz2010,Schulz2012}. Furthermore, the demonstration of the creation and annihilation of a single skyrmion\cite{Milde2013,Romming2013} shows their potential for the application in information technology. We measured a single-crystal Fe$_{0.5}$Co$_{0.5}$Si sample grown by optical float-zoning under ultra-high vacuum compatible conditions\cite{Neubauer2011}(courtesy of A. Bauer  and C. Pfleiderer, Technical University of Munich) in our dry cryostat observing (see figure \ref{fig:chiral_mag}(a)) helimagnetic structures at T = 3.2$\,$K in zero magnetic field and a skyrmion-lattice texture (see figure \ref{fig:chiral_mag}(b)) at T = 3.4$\,$K in an externally applied field B = 15$\,$mT.
The measurements of the helimagnetic phase in Fe$_{0.5}$Co$_{0.5}$Si and the imaging of the skyrmion-lattice phase transition of the single crystal was carried out using a sharp tip (tip apex radius $\sim$ 10$\,$nm) with magnetic coating, from Nanosensors (SSS-MFMR) to achieve the best lateral resolution. The magnetic tip was kept at a constant height of 20-30$\,$nm over the sample surface, with a phase-locked loop activated to monitor the cantilever resonance frequency. The sample was heated to 60$\,$K and the magnetic field was increased to 15$\,$mT. The sample was subsequently field-cooled to base temperature again. With the persistent switch heater of the superconducting magnet enabled, the temperature stabilized at 3.4$\,$K at the sample position and 3.6$\,$K at the magnet. MFM imaging may be carried out with sensors exhibiting increasingly higher force sensitivity, such as tuning forks\cite{Pelliccione2013}. In the next section we show shear-force tuning fork measurements performed in our dry cryostat which demonstrate the potential for further improvement, in particular towards the use of NV-centers in diamond-based scanning probe magnetometry\cite{Wrachtrup2008,Hong2013}.

\section{Tuning fork scanning force microscopy measurements}
\label{sec:tuningfork}
The implementation of quartz tuning forks as self-sensing probes in scanning force microscopy is especially beneficial because of their small size, the absence of thermal effects induced by the optical detection and the avoidance of unnecessary exposure of sensitive samples to light at cryogenic temperatures\cite{Karrai1995,Grober2000,Karrai2000}. To the best of our knowledge this is the first time that tuning-fork shear-force microscopy measurements are successfully reported in a dry cryostat.
\begin{figure}[ht]
\centering
\includegraphics[width=\columnwidth]{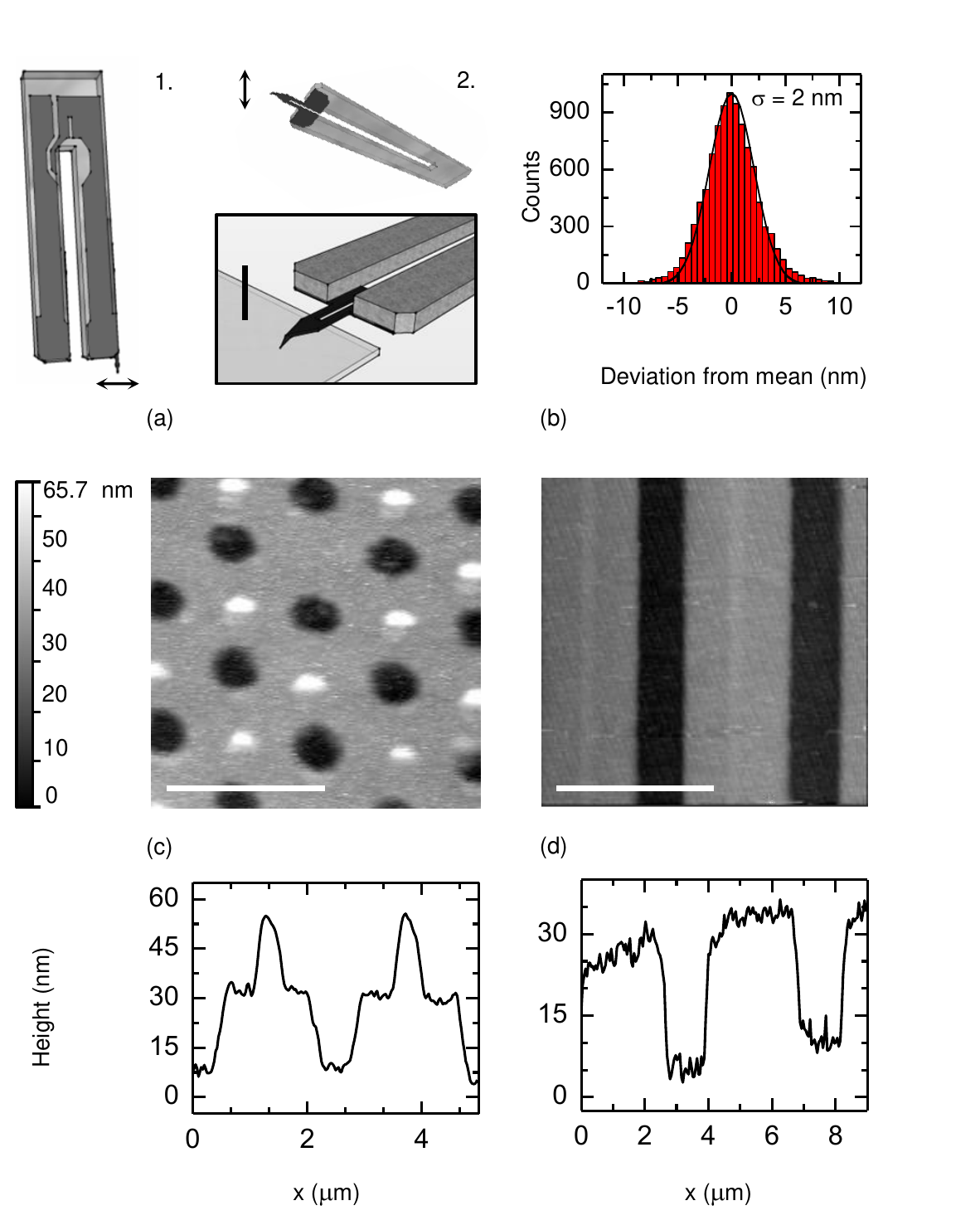}
\caption{(a) Quartz-crystal tuning fork with tungsten tip (1.) and with microcantilever (Akiyama\cite{Akiyama} probe, 2. and inset). The bar in the inset measures 0.2$\,$mm; (b) shear-force mode noise scan; the rms noise is 2$\,$nm measured with a bandwidth of 200$\,$Hz; (c) and (d) Quartz-crystal tuning fork atomic force microscopy scans of SiO$_{2}$ 20$\pm$2$\,$nm  high patterns on Si, performed in a dry cryostat. The scans were acquired at T = 4$\,$K with a resolution of 200 scan lines at 500 nm/s with two different scanning modes. Figure (c) shows a 5 $\times$ 5 $\,\mu m^{2}$ scan performed with an Akiyama\cite{Akiyama} probe (tapping mode; test grating pitch 2 $\mu m$). The bar in (c) measures 2$\,\mu m$. Figure (d) shows a 9 $\times$ 9 $\,\mu m^{2}$ scan performed with a quartz tuning fork coupled with a tungsten tip glued on one prong, operated in shear-force mode (test grating pitch 4$\,\mu m$). The bar in (d) measures 4$\,\mu m$. Cross sections relative to each scan are shown.}
\label{fig:tun_fork}
\end{figure}
A tungsten tip is etched\cite{Lucier2004} in house to reach a tip apex radius of approximately 80$\,$nm and glued to one prong of the tuning fork\cite{Karrai1995,Karrai2000}. The viscoelastic interaction of the oscillating fork takes place through shear forces between the sample and the tip and it can be roughly approximated according to the following expression
\begin{equation}
    F_{shear} = (1-\frac{V(z)}{V(z=\infty)})\frac{K\,x}{Q\,\sqrt{3}},
    \label{eq:frictionforce}
\end{equation}
where V(z) is the driving amplitude, $V(z=\infty)$ is the amplitude at the resonance of the free-oscillating fork, $K$ is the spring constant of the fork's prong, $x$ is the oscillation amplitude of the tuning fork, $Q$ is the quality factor of the resonator, which varies at different conditions of temperature, pressure, and distribution of mass across the resonator. A rough estimate of the magnitude of the shear forces may be obtained using $K \sim 30000\,$N/m, $Q = 3000$ at low temperature and a typical setpoint amplitude $V(z)=90\,\% V(z=\infty)$, which yields a value of $F \sim 50\,$pN assuming that the oscillation amplitude is kept as small as in the order of 0.1$\,$nm. Since typical amplitudes may be in the order of a few nanometers, a frictional interaction through shear forces in excess of 1-10$\,$nN causes deterioration of the image quality with reduction of the lateral resolution. Akiyama\cite{Akiyama} probes are quartz resonators where a sharp silicon microcantilever is placed at the ends of the prongs preserving the symmetry of the system. Typical values for the Akiyama spring constant are 5$\,$N/m. We demonstrate the achievement of tuning fork scanning force microscopy measurements of 20 $\pm 2\,$nm high SiO$_{2}$ patterns on Si with both Akiyama probes and in shear-force mode in our closed-cycle cryostat at T = 4$\,$K, as shown in figure \ref{fig:tun_fork}, in (c) and (d) respectively; the resolution is 200 lines per scan, acquired  at 500$\,$nm/s. The low noise amplitude  measured between the tuning fork and the sample with a bandwidth of 200 Hz and the feedback loop enabled shows a normal gaussian distribution with a standard deviation $\sigma$ = 2$\,$nm. Scanning in shear-force mode in a liquid-based cryostat, we measured noise amplitude better than 0.1$\,$nm rms and as discussed earlier, the contact mode noise measured in our dry cryostat is 65 pm. The reason for the excess noise here is that the residual noise in the system at the resonance of the tuning fork is considerably amplified from the resonator owing to its high quality factor. It would be possible to decouple the fork using dampers calibrated at its resonance frequency but for the time being this option is deemed not necessary. This study shows how our system has a ultra-low vibration level which enables satisfactory accomplishment of the most relevant applications of tuning fork in cryogenic environments, such as scanning gate microscopy or scanning magnetometry and other related magnetic resonance imaging techniques. Suspending the whole scanning system through springs would also represent a viable route to further reducing vibrations but this would defeat the purpose of the monolithic design which enables to combine free-space optical microscopy and scanning probe applications.
\section{Summary}
We have shown that the ultra-low vibration environment in our dry cryostat enables immediate application of scanning probe microscopy techniques. Contact-mode AFM measurements inside the dry cryostat resolved atomic steps of 0.39$\,$nm height in a SrTiO$_3$ sample, owing to a relative tip-sample vibration amplitude of less than 65$\,$pm. This level of noise allowed for the measurement of the elusive skyrmion-lattice phase in Fe$_{0.5}$Co$_{0.5}$Si by magnetic force microscopy. So far, this has only been achieved very recently in state-of-the-art liquid cryostat systems\cite{Milde2013}. Furthermore, quartz-crystal tuning fork shear-force microscopy has been achieved for the first time to the best of our knowledge in a dry cryostat. We studied the absolute vibrational noise in the sample space using a true inertial technique, measuring residual vibrations in the low nm range. We have also shown that these small displacements at very low frequencies do not influence the operation of the microscope and still enable measurements with high sensitivity. The salient feature of our development is that the microscope is top-loaded in the cryostat and does not require to be spring-suspended for further vibration isolation. This stiff architecture of the microscope head is of particular advantage when it comes to combine SPM with high resolution optical confocal microscopy where free-space optics is incompatible with spring mounting. The system as described allows a turnover of up to three experimental sessions per nine-hour working day. We hope that the techniques described in this paper will pave the way for a series of future measurements at low temperature, with the implementation of novel declinations of magnetic force microscopy, such as scanning diamond magnetometry\cite{Wrachtrup2008,Hong2013}.
\section*{Acknowledgements}
Two of us (F. P. Q. and J. P.) acknowledge financial support from Q-NET, a Marie Curie Initial Training Network funded by the European Commission (project number 264034). C. P. acknowledges financial support by the European Research Council (ERC AdG 291079, TOPFIT) and the German Science Foundation (TRR80, From Electronic Correlation to Functionality). A. B. acknowledges financial support through the TUM Graduate School.
\bibliography{C:/Users/Fran/Desktop/RSI/Bib}

\end{document}